\documentclass[prl,twocolumn,floatfix,a4paper,superscriptaddress]{revtex4}
\usepackage{bm,color,graphicx,amsmath,txfonts}

\begin{document}

\title{Squeezed states of magnons and phonons in cavity magnomechanics}

\author{Jie Li}
\affiliation{Department of Physics, Zhejiang University, Hangzhou 310027, China}
\affiliation{Institute for Quantum Science and Engineering and Department of Biological and Agricultural Engineering, Texas A{\rm \&}M University, College Station, Texas 77843, USA}
\author{Shi-Yao Zhu}
\affiliation{Department of Physics, Zhejiang University, Hangzhou 310027, China}
\author{G. S. Agarwal}
\affiliation{Institute for Quantum Science and Engineering and Department of Biological and Agricultural Engineering, Texas A{\rm \&}M University, College Station, Texas 77843, USA}
\affiliation{Department of Physics and Astronomy, Texas A{\rm \&}M University, College Station, Texas 77843, USA}

\begin{abstract}
We show how to create quantum squeezed states of magnons and phonons in a cavity magnomechanical system. The magnons are embodied by a collective motion of a large number of spins in a macroscopic ferrimagnet, and couple to cavity microwave photons and phonons (vibrational modes of the ferrimagnet) via the magnetic dipole interaction and magnetostrictive interaction, respectively. The cavity is driven by a weak squeezed vacuum field generated by a flux-driven Josephson parametric amplifier, which is essential to get squeezed states of the magnons and phonons. We show that the magnons can be prepared in a squeezed state via the cavity-magnon beamsplitter interaction, and by further driving the magnon mode with a strong red-detuned microwave field, the phonons are squeezed. We show optimal parameter regimes for obtaining large squeezing of the magnons and phonons, which are robust against temperature and could be realized with experimentally reachable parameters. 
\end{abstract}

\date{\today}
\maketitle

{\it Introduction.} Recent years have witnessed a significant progress of realizing strong light-matter interaction using collective excitations of spin ensembles in ferrimagnetic systems, for example in yttrium iron garnet (YIG), thanks to their very high spin density and low damping rate. The Kittel mode~\cite{Kittel} (uniformly precessing mode) in the YIG sphere can strongly couple to the microwave cavity photons leading to cavity polaritons~\cite{Strong1,Strong2,Strong3,Strong4,Strong5,Strong6}. Many other interesting phenomena have been studied in the system of cavity-magnon polaritons, such as the observation of magnon dark modes~\cite{TangNC}, the exceptional point~\cite{YouNC}, manipulation of distant spin currents~\cite{spinCur}, and bistability~\cite{You18}. Another stimulating direction is to extend cavity-magnon systems to hybrid systems by coupling the magnons to a superconducting qubit~\cite{Naka15}, or phonons~\cite{Tang16}, which allow one to resolve magnon number states~\cite{Naka17}, or generate tripartite entangled states of magnons, photons, and phonons~\cite{JiePRL}.

Magnon systems, in particular the YIG spheres, provide a totally new platform for the study of macroscopic quantum phenomena. It is natural to study macroscopic quantum states in magnon-photon systems owing to a large size of the YIG sphere. An important quantum state would be a squeezed state. Squeezed states can be used not only to improve the measurement sensitivity~\cite{Caves}, but also to study decoherence theories at large scales~\cite{collapse}. Besides, squeezed states are a vital ingredient for continuous variable information processing~\cite{Loock}. Furthermore, the magnon-phonon interactions in YIG spheres are well studied. The nonlinear magnetostrictive (radiation pressure-like) interaction allows one to create magnomechanical entanglement which then transfers to the photon-magnon and photon-phonon subsystems, forming a tripartite entangled state~\cite{JiePRL}. Such an interaction can also be used to cool the mechanical motion and to generate a variety of nonclassical states in both the magnon and mechanical modes by suitably driving the YIG sphere. We note that the field of cavity optomechanics~\cite{OMRMP} has witnessed a considerable progress in observing quantum effects in massive systems, where quantum squeezing of mechanical motion~\cite{Schwab}, nonclassical correlations between single photons and phonons~\cite{Simon16}, and quantum entanglement between two massive mechanical oscillators~\cite{enMM1,enMM2} have been observed.

Here we present a scheme to generate squeezed states of both the magnons and phonons in a hybrid magnon-photon-phonon system. Typically the generation of squeezing is based on the nonlinearity of the system, for example Kerr nonlinearity in the case of optical fibers~\cite{Kumar}. For magnons the nonlinearity is small and hence we follow an alternative method where we drive the cavity with a squeezed field~\cite{Zoller,GA09}. The microwave cavity is driven by a weak squeezed vacuum field generated by a flux-driven Josephson parametric amplifier (JPA), which is used to shape the noise properties of quantum fluctuations of the cavity field, leading to a squeezed cavity field. The squeezing is then transferred to the magnons due to the effective cavity-magnon beamsplitter interaction. By further driving the magnon mode with a strong red-detuned microwave field, the magnomechanical state-swap interaction is activated, resulting in the transferring of squeezing from the magnons to phonons. The generated squeezed states are in the steady state and robust against environmental temperature. We study the system by using the standard Langevin formalism and by linearizing the dynamics, and we finally discuss the validity of our linearized model and provide strategies to measure the squeezing.

\begin{figure}[t]
\hskip-0.08cm\includegraphics[width=\linewidth]{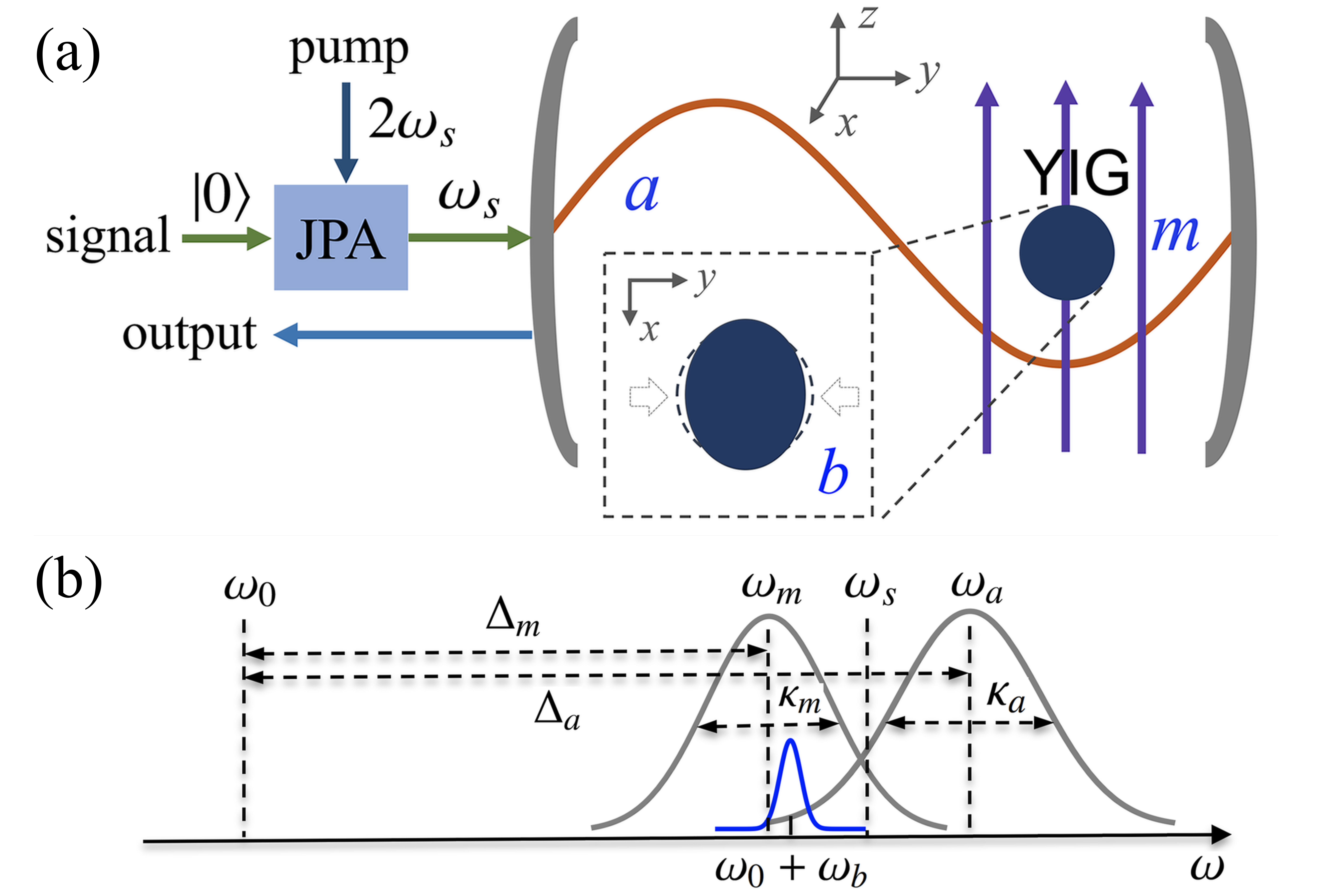} 
\caption{(a) A YIG sphere is simultaneously in a uniform bias magnetic field and near the maximum magnetic field of a microwave cavity mode, which is driven by a weak squeezed vacuum field generated by a flux-driven JPA. The magnon mode is directly driven by a microwave source (not shown) to enhance the magnomechanical coupling. The bias magnetic field ($z$ direction), the drive magnetic field ($y$ direction) and the magnetic field ($x$ direction) of the cavity mode are mutually perpendicular at the site of the YIG sphere. (b) Frequencies and linewidths of the system. The magnon mode with frequency $\omega_m$ and linewidth $\kappa_m$ is driven by a microwave field at frequency $\omega_0$ and the mechanical motion at frequency $\omega_b$ scatters photons onto the two sidebands at $\omega_0 + \omega_b$ (blue sideband) and $\omega_0 - \omega_b$ (red sideband, not shown). If the drive squeezed field with frequency $\omega_s$, the cavity with frequency $\omega_a$ and linewidth $\kappa_a$, and the magnon mode are resonant, the magnon mode is squeezed, and if they are further resonant with the blue (anti-Stokes) sideband, the mechanical mode is cooled and squeezed. }
\label{fig1}
\end{figure}

{\it The model}. We consider a cavity magnomechanical system~\cite{Tang16,JiePRL}, as shown in Fig.~\ref{fig1} (a), which consists of cavity microwave photons, magnons, and phonons. The cavity is one sided and the output field of the cavity can be used for measuring the magnon state. The magnons are embodied by a collective motion of a large number of spins in a YIG sphere, and couple to the cavity photons via the magnetic dipole interaction and to the phonons via the magnetostrictive interaction~\cite{Kittel2}. %Specifically, the varying magnetization induced by the magnon excitation inside the YIG sphere leads to the deformation of its geometry structure, which forms vibrational modes (phonons) of the sphere, and vice versa~\cite{Kittel2}. 
We consider the size of the sphere to be much smaller than the microwave wavelength, and the Hamiltonian of the system reads
\begin{equation}
\begin{split}
{\cal H}/\hbar &= \omega_a a^{\dag} a + \omega_m m^{\dag} m + \frac{\omega_b}{2} (q^2 + p^2) + g_{mb} m^{\dag} m q   \\
&+ g_{ma} (a + a^{\dag}) (m + m^{\dag}) + i \Omega (m^{\dag} e^{-i \omega_0 t}  - m e^{i \omega_0 t} ),
\end{split}
\end{equation}
where $a$ and $m$ ($a^{\dag}$ and $m^{\dag}$) are the annihilation (creation) operator of the cavity and magnon modes, respectively, $[O, O^{\dag}]\,{=}\,1$ ($O\,{=}\,a,m$), $q$ and $p$ ($[q, p]\,{=}\,i$) are the dimensionless position and momentum quadratures of the mechanical mode, and $ \omega_a$, $ \omega_m$, and $ \omega_b$ are the resonance frequencies of the cavity, magnon and mechanical modes, respectively. The magnon frequency is determined by the external bias magnetic field $H$ via $\omega_m=\gamma H$, where $\gamma/2\pi= 28$ GHz/T is the gyromagnetic ratio. The magnon-cavity coupling rate $g_{ma}$ currently can be (much) larger than the dissipation rates, $\kappa_m$ and $\kappa_a$, of the magnon and cavity modes, such that the strong coupling regime $g_{ma} > \kappa_{m}, \kappa_{a}$ can be achieved~\cite{Strong1,Strong2,Strong3,Strong4,Strong5}. The bare magnomechanical coupling rate $g_{mb}$ is currently small, but the magnomechanical interaction can be efficiently enhanced by directly driving the magnon mode with a strong microwave field~\cite{You18,You16}. The Rabi frequency $\Omega =\frac{\sqrt{5}}{4} \gamma \! \sqrt{N} B_0$~\cite{JiePRL} denotes the coupling strength of the drive magnetic field (with amplitude $B_0$ and frequency $\omega_0$) with the magnon mode, where the total number of spins $N=\rho V$ with $\rho=4.22 \times 10^{27}$ m$^{-3}$ the spin density of the YIG and $V$ the volume of the sphere. It should be noted that the boson operators $m$ and $m^{\dag}$ describe the collective motion of the spins via the Holstein-Primakoff transformation~\cite{HPT}, and $\Omega$ is derived under the condition of the low-lying excitations, $\langle m^{\dag} m \rangle \ll 2Ns$, where $s=\frac{5}{2}$ is the spin number of the ground state Fe$^{3+}$ ion in YIG.

We apply the rotating-wave approximation to the magnon-photon interaction, $g_{ma} (a + a^{\dag}) (m + m^{\dag}) \to g_{ma} (a m^{\dag} + a^{\dag} m)$~\cite{Strong1,Strong2,Strong3,Strong4,Strong5,Tang16}, and due to this interaction the squeezing from the cavity mode is transferred to the magnon mode. This interaction is typically referred to as the beamsplitter interaction. In the frame rotating at the drive frequency $\omega_0$, the quantum Langevin equations (QLEs) describing the system are as follows
\begin{equation}\label{QLE1}
\begin{split}
\dot{a}&= - (i \Delta_a + \kappa_a) a - i g_{ma} m + \sqrt{2 \kappa_a} a^{\rm in},  \\
\dot{m}&= - (i \Delta_m + \kappa_m) m - i g_{ma} a - i g_{mb} m q + \Omega + \sqrt{2 \kappa_m} m^{\rm in},  \\
\dot{q}&= \omega_b p,   \\
\dot{p}&= - \omega_b q - \gamma_b p - g_{mb} m^{\dag}m + \xi, 
\end{split}
\end{equation}
where $\Delta_{a} \,\,{=}\,\, \omega_{a} \,\,{-}\,\, \omega_0$, $\Delta_{m} \,\,{=}\,\, \omega_{m} \,\,{-}\,\, \omega_0$, $\gamma_b$ is the mechanical damping rate, and $a^{\rm in}$, $m^{\rm in}$ and $\xi$ are input noise operators for the cavity, magnon and mechanical modes, respectively, which are zero mean and characterized by the following correlation functions~\cite{Gardiner}: $\langle a^{\rm in}(t) \, a^{\rm in \dag}(t')\rangle = ({\cal N}{+}1) \,\delta(t{-}t')$, $\langle a^{\rm in \dag}(t) \, a^{\rm in}(t')\rangle \,\,{=}\,\, {\cal N} \, \delta(t{-}t')$, $\langle a^{\rm in}(t) \, a^{\rm in}(t')\rangle {=} {\cal M} \, e^{-i \Delta_s (t+t')} \delta(t{-}t')$, and $\langle a^{\rm in \dag}(t) \, a^{\rm in \dag}(t')\rangle = {\cal M}^* \,\, e^{i \Delta_s (t+t')}\, \delta(t{-}t')$, where ${\cal N}=\sinh^2 r$, ${\cal M}=e^{i \theta} \sinh r \cosh r $, and $\Delta_s\,{=}\,\omega_s \,{-}\, \omega_0$, with $r$, $\theta$, and $\omega_s$ being respectively the squeezing parameter, phase, and frequency of the squeezed vacuum field, which is generated by degenerate parametric down-conversion in a flux-driven JPA~\cite{JPA1,JPA2,JPA3,JPA4,JPA5,JPA6,JPA7,JPA8,JPA9,JPA10,JPA11} based on the nonlinearity of Josephson junctions. It has been reported that the degree of squeezing as high as 10 dB has been produced~\cite{JPA3}. If the JPA is working with a pump field at frequency $2\omega_s$ and vacuum fluctuations at the signal input port, it generates a squeezed vacuum microwave field at frequency $\omega_s$ [see Fig.~\ref{fig1} (a)]~\cite{JPA4,JPA7}. Alternatively, one can produce a squeezed vacuum field by a degenerate parametric amplifier with a microwave pump~\cite{Siddiqi}. The other input-noise correlation functions are $\langle m^{\rm in}(t) \, m^{\rm in \dag}(t')\rangle = [N_m(\omega_m)+1] \, \delta(t{-}t')$, $\langle m^{\rm in \dag}(t) \, m^{\rm in}(t')\rangle = N_m(\omega_m)\, \delta(t{-}t')$, and $\langle \xi(t)\xi(t')\,{+}\,\xi(t') \xi(t) \rangle/2 \,\, {\simeq} \,\, \gamma_b [2 N_b(\omega_b) {+}1] \delta(t{-}t')$, where a Markovian approximation has been made, which is valid for a large mechanical quality factor ${\cal Q}= \omega_b/\gamma_b \,\, {\gg}\, 1$~\cite{Markov}, and $N_j(\omega_j){=}\big[ {\rm exp}\big( \frac{\hbar \omega_j}{k_B T} \big) {-}1 \big]^{-1} $ $(j{=}m,b)$ are the equilibrium mean thermal magnon and phonon numbers, respectively.

\begin{figure}[t]
\hskip-0.15cm\includegraphics[width=0.95\linewidth]{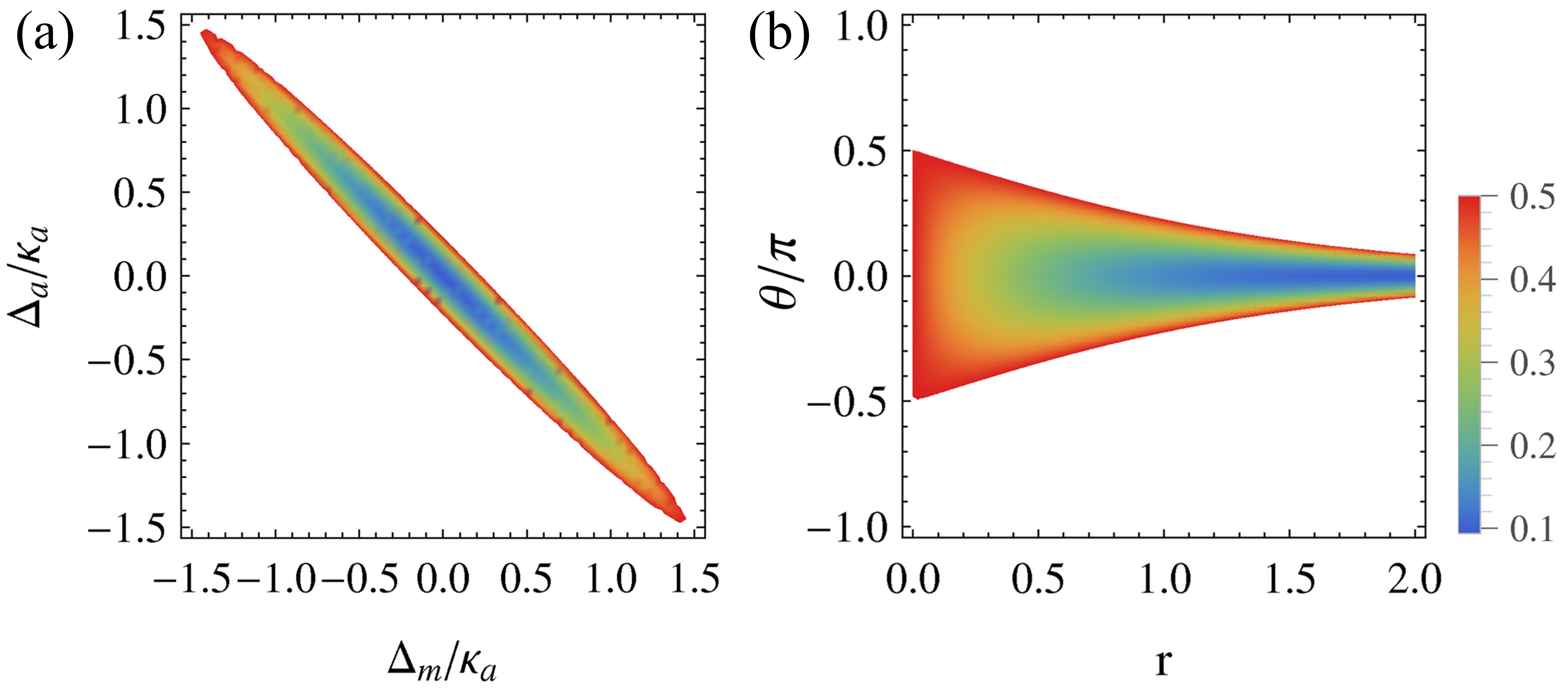} 
\caption{Variance of the magnon amplitude quadrature $\langle \delta x(t)^2  \rangle$ versus (a) detunings $\Delta_m$ and $\Delta_a$, and (b) squeezing parameter $r$ and phase $\theta$. In (a) $r=2$, and $\theta=0$, and in (b) $\Delta_a=\Delta_m=0$.  The blank area denotes $\langle \delta x(t)^2  \rangle >\frac{1}{2}$, i.e., above vacuum fluctuations. See text for details of the other parameters.}
\label{fig2}
\end{figure}

\begin{figure}[t]
\hskip-0.08cm\includegraphics[width=\linewidth]{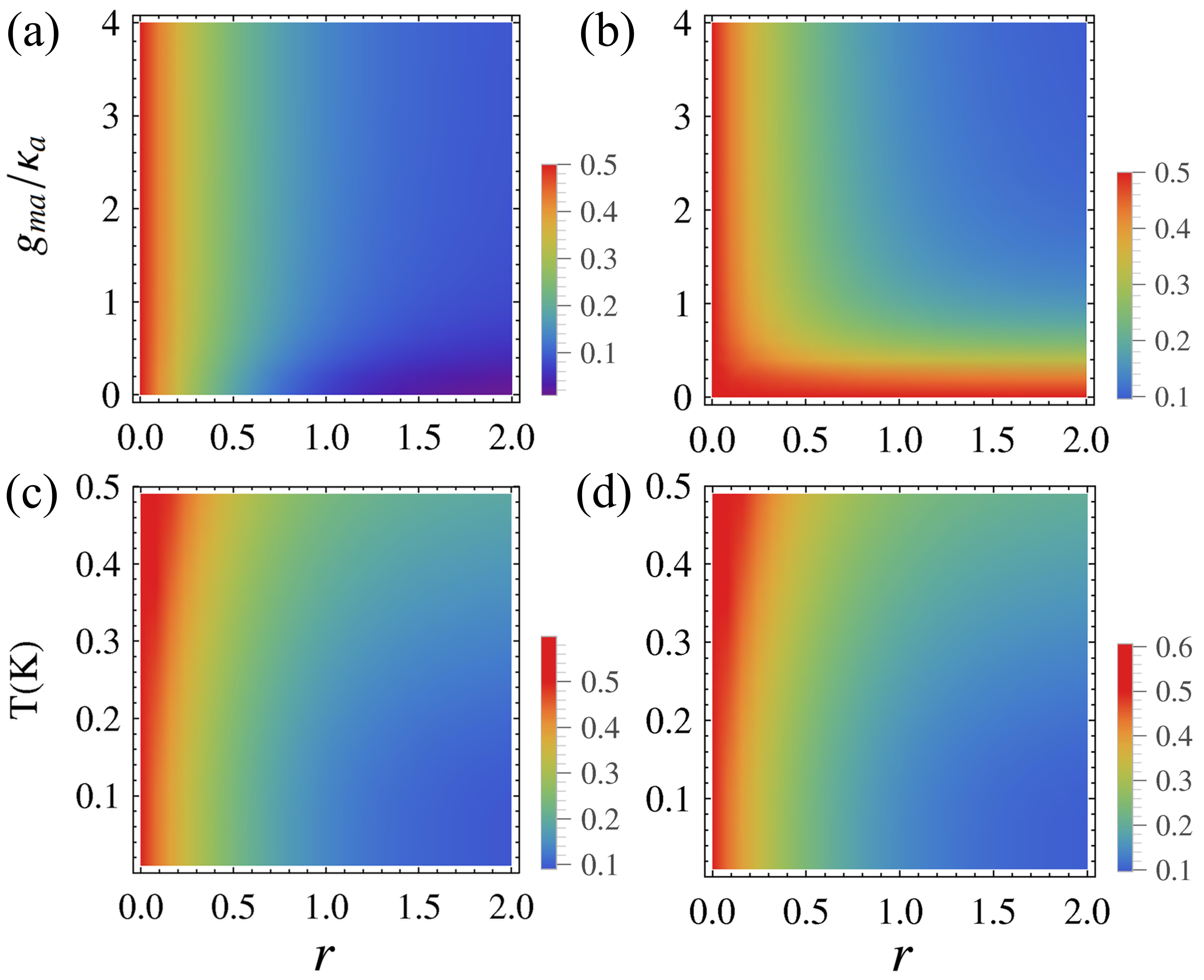} 
\caption{(a) Variance of the cavity phase quadrature $\langle \delta Y(t)^2  \rangle$ and (b) of the magnon amplitude quadrature $\langle \delta x(t)^2  \rangle$ versus squeezing parameter $r$ and coupling $g_{ma}$. (c) $\langle \delta Y(t)^2  \rangle$ and (d) $\langle \delta x(t)^2  \rangle$ versus $r$ and temperature $T$ (from 10 mK to 500 mK). We take $T=20$ mK for (a) and (b), and $g_{ma}=4 \kappa_a$ for (c) and (d), and $\Delta_a=\Delta_m=0$ and $\theta=0$ for all the plots. The other parameters are as in Fig.~\ref{fig2}. }
\label{fig3}
\end{figure}

{\it Squeezing the magnons.} We first show that the magnons can be squeezed by resonantly driving the cavity with a squeezed vacuum field, and then in the next section we show that this squeezing can be transferred to the phonons by further driving the magnon mode with a strong red-detuned field. In the absence of the magnon drive and due to the fact that $g_{mb} \ll g_{ma}$~\cite{Note}, the mechanical mode is {\it effectively} decoupled with the magnon mode, and the system then reduces to a two-mode system with zero averages. The fluctuations of the system are described by the QLEs
\begin{equation}\label{QLE2}
\begin{split}
\delta\dot{a}&= - (i \Delta_a + \kappa_a) \delta a - i g_{ma} \delta m + \sqrt{2 \kappa_a} a^{\rm in},  \\
\delta\dot{m}&= - (i \Delta_m + \kappa_m) \delta m - i g_{ma} \delta a + \sqrt{2 \kappa_m} m^{\rm in},  \\
\end{split}
\end{equation}
which are linear and can be solved straightforwardly~\cite{SM}. Here $\Delta_a$ and $\Delta_m$ are redefined with respect to the drive frequency $\omega_s$. For the resonant case $\Delta_a = \Delta_m =0$, we obtain relatively simple expressions for the variances of the {\it squeezed} cavity and magnon quadratures, $\delta Y=i(\delta a^{\dag} - \delta a)/\sqrt{2}$, and $\delta x=(\delta m + \delta m^{\dag})/\sqrt{2}$, which are respectively
\begin{equation}\label{SYSx}
\begin{split}
&\langle \delta Y(t)^2  \rangle  \\
&{=} \frac{ g_{ma}^2 (2N_m{+}1) \kappa_m {+}\kappa_a (g_{ma}^2 {+} \kappa_a \kappa_m {+} \kappa_m^2 )  (\cosh 2r {-} \cos \theta \sinh 2r )     }{2(\kappa_a+\kappa_m) (g_{ma}^2+\kappa_a \kappa_m) },   \\
&\langle \delta x(t)^2  \rangle   \\
&{=}  \frac{  (2N_m{+}1) \kappa_m (g_{ma}^2 {+} \kappa_a \kappa_m {+} \kappa_a^2 ) + g_{ma}^2 \kappa_a (\cosh 2r {-} \cos \theta \sinh 2r )     }{2(\kappa_a+\kappa_m) (g_{ma}^2+\kappa_a \kappa_m) }. 
\end{split}
\end{equation}
The variance of the magnon amplitude quadrature $\langle \delta x(t)^2  \rangle$ takes a simple form 
\begin{equation}
\langle \delta x(t)^2  \rangle \simeq \frac{1}{2} (e^{-2r} + \frac{\kappa_m}{\kappa_a}),
\end{equation}
for the optimal squeezing regime: $\theta\,{=}\,0$, $g_{ma} \,{\gg}\, \kappa_a \,{\gg}\, \kappa_m$ ($\kappa_a \gg \kappa_m$ for the detection of the magnon state~\cite{JiePRL}), and at low temperature $N_m \simeq 0$, where the magnon and cavity modes achieve the same degree of squeezing. Such an optimal regime is confirmed by the results of Figs.~\ref{fig2} and~\ref{fig3}. In our definition $\langle \delta Q(t)^2  \rangle =\frac{1}{2}$ ($Q$ is a mode quadrature) denotes vacuum fluctuations. The degree of squeezing can be expressed in the dB unit, which can be evaluated by $-10\, {\rm log}_{10} \big[ \langle \delta {Q}(t)^2  \rangle/\langle \delta {Q}(t)^2  \rangle_{\rm vac} \big]$, where $\langle \delta {Q}(t)^2  \rangle_{\rm vac}=\frac{1}{2}$.

Figure~\ref{fig2} shows that $\Delta_a=\Delta_m=0$ (i.e., $\omega_a = \omega_m= \omega_s$), and the phase $\theta=0$ are optimal for magnon squeezing. We have adopted experimentally feasible parameters~\cite{Strong2}: $\omega_a/2\pi=10$ GHz, $\kappa_a/2\pi \,{=}\, 5\kappa_m/2\pi \,{=}\, 5$ MHz, $g_{ma} \, {=} \,4 \kappa_a$, and $T\,{=}\,20$ mK, where we have used a larger $\kappa_a$ for reading out the magnon state~\cite{JiePRL}. Figure~\ref{fig3} (a) and (b) show clearly the process of the squeezing transferring from the cavity to the magnon mode. When the cavity photons and magnons are decoupled ($g_{ma}\,{=}\,0$), the cavity field is squeezed as a result of the squeezed driving field, while the magnon mode possesses vacuum fluctuations as $N_m \simeq 0$ at 20 mK. As the coupling grows, the cavity squeezing decreases while the magnon squeezing increases, implying that the squeezing has been partially transferred from the cavity to the magnons. For example, for $r\,{=}\,1$ the cavity squeezing is about 8.69 dB when $g_{ma}\,{=}\,0$, and the magnon (cavity) squeezing is about 5.40 dB (5.56 dB) when $g_{ma}/2\pi\,{=}\,20$ MHz (In Ref.~\cite{Strong2} $g_{ma}/2\pi\,{=}\,47$ MHz has been realized). Note that we have assumed the bandwidth of the squeezed vacuum field is larger than the cavity linewidth. In Ref.~\cite{JPA11} a 3.87 dB squeezed vacuum with a bandwidth of 30 MHz has been produced. This will yield a 2.89 dB squeezing of the magnons for $g_{ma}/2\pi\,{=}\,20$ MHz. The cavity and magnon squeezing are quite robust against the temperature, as shown in Fig.~\ref{fig3} (c) and (d). Moderate squeezing in both the cavity and magnon quadratures can be found even at $T{=}0.5$ K. The squeezed microwave drive produces squeezing of the collective mode in steady state. It, however, does not affect the decay of the mode, which is in contrast to the behavior of a spin in squeezed field~\cite{Siddiqi,Gardiner}. Thus we study the amount of squeezing in steady state.

%We are aware that squeezed light can affect the dynamical behavior of a spin~\cite{Gardiner} as observed in the experiment~\cite{Siddiqi}. Here we study the steady-state behavior. The spin system has a behavior which is different from that exhibited by a collective mode. In contrast, the collective mode's dynamics is not affected by the squeezed bath, but it gets squeezed in the steady state.

{\it Squeezing the phonons.} Once the magnons are squeezed, we turn on the strong red-detuned magnon drive [see Fig.~\ref{fig1} (b)] to activate the magnomechanical state-swap operation. We adopt the linearization treatment as used in Ref.~\cite{JiePRL}, and the linearized QLEs describing the system quadrature fluctuations $(\delta X, \delta Y, \delta x, \delta y, \delta q, \delta p)$, where $\delta X=(\delta a + \delta a^{\dag})/\sqrt{2}$, and $\delta y=i(\delta m^{\dag} - \delta m)/\sqrt{2}$, can be cast in the matrix form
\begin{equation}\label{uAn}
\dot{u} (t) = A u(t) + n(t) ,
\end{equation}
where $u(t)=\big[\delta X (t), \delta Y (t), \delta x (t), \delta y (t), \delta q (t), \delta p (t) \big]^T$, $n (t) = \big[ \!\sqrt{2\kappa_a} X^{\rm in} (t), \sqrt{2\kappa_a} Y^{\rm in} (t), \sqrt{2\kappa_m} x^{\rm in} (t), \sqrt{2\kappa_m} y^{\rm in} (t), 0, \xi (t) \big]^T$ is the vector of input noises, and the drift matrix $A$ is given by  
\begin{equation}\label{AAA}
A =
\begin{pmatrix}
-\kappa_a  &  \Delta_a  &  0 &  g_{ma}  &  0  &  0   \\
-\Delta_a  & -\kappa_a  & -g_{ma}  & 0  &  0  &  0   \\
0 & g_{ma}  & -\kappa_m  & \tilde{ \Delta}_m &  -G_{mb}  &  0 \\
-g_{ma}  & 0 & -\tilde{ \Delta}_m & -\kappa_m &  0  &  0 \\
0 &  0  &  0  &  0  &  0  &  \omega_b   \\
0 &  0  &  0  &  G_{mb}  & -\omega_b & -\gamma_b   \\
\end{pmatrix} ,
\end{equation}
where $\tilde{ \Delta}_m = \Delta_m + g_{mb} \langle q \rangle$ is the effective magnon-drive detuning including the frequency shift due to the magnon-phonon interaction, and $G_{mb} = i \sqrt{2} g_{mb} \langle m \rangle$ is the effective magnomechanical coupling rate, where $\langle q \rangle = - \frac{g_{mb}}{\omega_b} |\langle m \rangle|^2 $, and
\begin{equation}\label{eq5}
\langle m \rangle =  \frac{  \Omega  ( i \Delta_a +\kappa_a) }{ g_{ma}^2 \! + ( i \tilde{ \Delta}_m + \kappa_m) ( i \Delta_a + \kappa_a) },
\end{equation}
which becomes $\langle m \rangle \simeq  i  \Omega  \Delta_a/ (g_{ma}^2 \, {-} \, \tilde{ \Delta}_m \Delta_a ) $ when $|\tilde{ \Delta}_m|, |\Delta_a| \gg  \kappa_a, \kappa_m$, which is a pure imaginary number. The drift matrix $A$ is given under this condition. %In fact, we will show later that $\tilde{ \Delta}_m \simeq \Delta_a \simeq \omega_b  \gg  \kappa_a, \kappa_m$ [see Fig.~\ref{fig1} (b)] are optimal for the mechanical squeezing. 
By taking the Fourier transform of Eq.~\eqref{uAn} and solving it in the frequency domain, the variances of the mechanical position and momentum quadratures $\langle \delta q(t)^2  \rangle$ and $\langle \delta p(t)^2  \rangle$ can be obtained~\cite{SM}, which are, however, too lengthy to be reported here.

\begin{figure}[t]
\hskip-0.08cm\includegraphics[width=\linewidth]{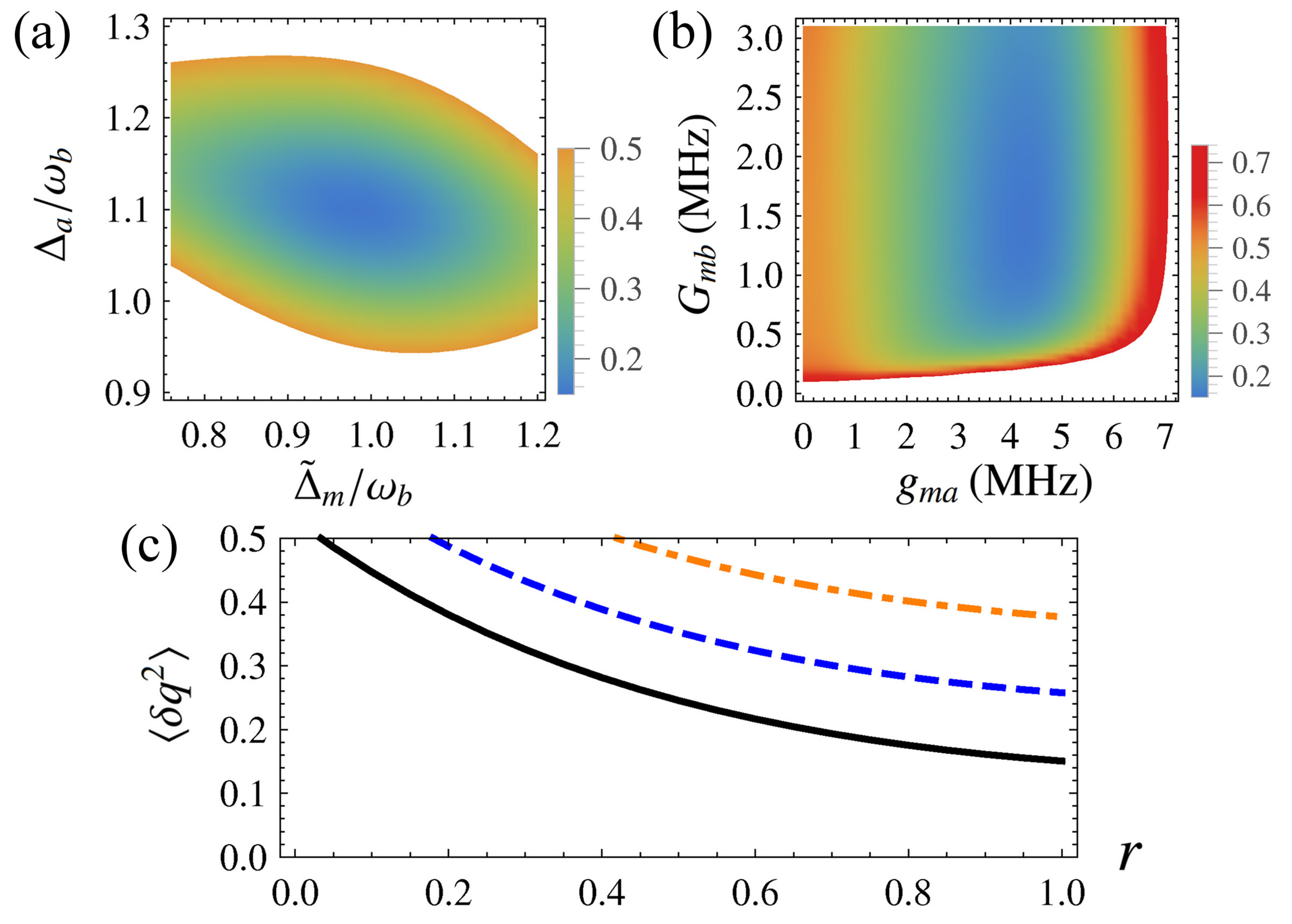} 
\caption{Variance of the mechanical position quadrature $\langle \delta q(t)^2  \rangle$ versus (a) detunings $\tilde{ \Delta}_m$ and $\Delta_a$, (b) couplings $g_{ma}$ and $G_{mb}$, and (c) squeezing parameter $r$ for temperatures $T=10$ mK (solid), 100 mK (dashed), and 200 mK (dot-dashed). We take $g_{ma}/2\pi= 4.2$ MHz and $G_{mb}/2\pi= 1.5$ MHz in (a) and (c), $\tilde{ \Delta}_m = \omega_b$ and $\Delta_a = 1.1\omega_b$ in (b)-(c), $r=1$ in (a)-(b), and $\Delta_s =\omega_b$ and $\theta=0$ for all the plots. In (a) [(b)] the blank area denotes $\langle \delta q(t)^2  \rangle >0.5$ (0.74). See text for details of the other parameters. }
\label{fig4}
\end{figure}

Figure~\ref{fig4} shows the variance of the mechanical position quadrature $\langle \delta q(t)^2  \rangle$ versus various key parameters of the system. All the results are in the steady state guaranteed by the negative eigenvalues (real parts) of the drift matrix $A$. We have employed experimentally feasible parameters~\cite{Tang16,Note3}: $\omega_a/2\pi \,{=}\,10$ GHz, $\omega_b/2\pi \,{=}\, 10$ MHz, $\gamma_b/2\pi \,{=}\,10^2$ Hz, $\kappa_a/2\pi=5\kappa_m/2\pi=3$ MHz, and at low temperature $T\,{=}\,10$ mK. Clearly, there is an optimal regime around $\tilde{ \Delta}_m \simeq \Delta_a \simeq \omega_b$ [see Fig.~\ref{fig4} (a)], where the mechanical squeezing is prominent. At this detuning, %a drive microwave photon with frequency $\omega_m-\omega_b$ interacts with a phonon with frequency $\omega_b$ and they convert into a magnon with frequency $\omega_m$, 
the effective Hamiltonian of the magnomechanical interaction is $H_{\rm eff}/\hbar \,{=}\, G_{mb}(m^{\dag} b + m b^{\dag})$~\cite{Jie18}, where $b=(q+ip)/\sqrt{2}$, which results in the mechanical mode significantly cooled close to the ground state. This interaction also realizes the magnon-phonon state-swap operation leading to the transferring of squeezing to the phonons. Figure~\ref{fig4} (b) shows that an optimal coupling $g_{ma}$ exists as a result of the balance between the efficiency of squeezing transferring and the heating of the mechanical mode. For optimal detunings, the effective magnomechanical coupling $G_{mb} \, {\simeq} \sqrt{2} g_{mb} \frac{\omega_b \Omega}{\omega_b^2 - g_{ma}^2}$, and $G_{mb}$ takes moderate values keeping the nonlinear effect negligible (we verify this later on). The optimal coupling $G_{mb}/2\pi =1.5$ MHz in Fig.~\ref{fig4} corresponds to the drive magnetic field $B_0 \simeq 3.0 \times 10^{-5}$ T and the drive power $P \simeq 5.3$ mW~\cite{B0P} for $g_{mb}/2\pi = 0.1$ Hz and an optimal coupling $g_{ma}/2\pi \simeq 4.2$ MHz, which yield the largest mechanical squeezing of 5.21 dB for the squeezed vacuum driving field of $r=1$ (8.69 dB). The squeezing is robust against temperature as shown in Fig.~\ref{fig4} (c), which is still below vacuum fluctuations when $T=200$ mK and $r>0.42$ (3.65 dB). For a 3.87 dB squeezed vacuum drive~\cite{JPA11}, a mechanical squeezing of 2.77 dB (2.05 dB) can be achieved at 10 mK (50 mK).

The above results are valid only when the assumption of the low-lying excitations, $\langle m^{\dag} m \rangle \ll 2Ns=5N$, is satisfied. For what we use a 250-$\mu$m-diameter YIG sphere~\cite{Tang16}, the number of spins $N \,{\simeq}\, 3.5 \times 10^{16}$, and $G_{mb}/2\pi \,{=}\,1.5$ MHz corresponds to $|\langle m \rangle| \,{\simeq}\, 1.1 \times 10^7$ and $\Omega \,{\simeq}\, 5.5 \times 10^{14}$ Hz. Therefore, $\langle m^{\dag} m \rangle \, {\simeq} \, 1.1 \times 10^{14} \,{\ll}\, 5N \,{=}\,1.7 \times 10^{17}$, and the assumption is thus well satisfied. The strong magnon drive may bring about unwanted nonlinear effects due to the Kerr nonlinear term ${\cal K} m^{\dag}m m^{\dag} m$ in the Hamiltonian~\cite{You18,You16}, with ${\cal K}$ the Kerr coefficient, which is inversely proportional to the volume of the sphere. For a 1-mm-diameter YIG sphere used in Refs.~\cite{You18,You16}, ${\cal K}/2\pi \,\,{\approx}\,\, 10^{-10}$ Hz, and thus for a 250-$\mu$m-diameter sphere, ${\cal K}/2\pi \,\,{\approx}\,\, 6.4 \times 10^{-9}$ Hz. To have a negligible nonlinear effect, ${\cal K}|\langle m \rangle|^3 \ll \Omega$ must hold. For the optimal coupling $G_{mb}/2\pi \,{=}\, 1.5$ MHz used in Fig.~\ref{fig4}, we have ${\cal K}|\langle m \rangle|^3 \simeq 4.8 \times 10^{13}$ Hz $\ll \Omega \simeq 5.5 \times 10^{14}$ Hz, which implies that our linearized model is valid and a good approximation.

Finally, we discuss how to detect the magnon and mechanical squeezing. The magnon state can be read out by sending a weak microwave probe field and by homodyning the cavity output of the probe field. This requires that the cavity decay rate should be much larger than the magnon dissipation rate, such that when the drive is switched off and all cavity photons decay the magnon state remains effectively unchanged, at which time a probe field is sent. Another possibility is to study the changes in the squeezing spectrum of the cavity output, which carries a clear signature of magnon squeezing~\cite{SM}. The mechanical squeezing can be measured by coupling the YIG sphere to an additional optical cavity which is driven by a weak red-detuned light. In this case, the squeezed phonon state is mapped onto the squeezed cavity output field~\cite{Jie18,Hofer}. Note that, to see the squeezing of the phonon mode, it is preferable to use the optical cavity as then the radiation pressure interaction is quite prominent and the squeezing of phonons is effectively transferred to the squeezing of the output.

{\it Conclusions}. We have presented a scheme to generate squeezed states of magnons and phonons in a hybrid cavity magnomechanical system. By driving the cavity with a squeezed vacuum field and activating the effective cavity-magnon and magnon-phonon state-swap interactions, the magnons and phonons are squeezed in succession. Moderate squeezing of the magnons (phonons) can be achieved by using experimentally feasible (reachable) parameters. Larger squeezing could be realized by increasing the degree of squeezing of the drive field and working at a lower temperature. The squeezed states of the magnons and phonons realized in a massive object represent genuinely macroscopic quantum states~\cite{note2}, and are thus useful in the study of quantum-to-classical transition, test of decoherence theories~\cite{collapse}, as well as for ultrasensitive detections. The hybrid system may find its applications in quantum information processing, where the mechanical oscillator can act as storage of information which can be transferred to photonic and magnonic systems.

{\it Acknowledgments}. This work has been supported by the National Key Research and Development Program of China (Grants No. 2017YFA0304200, and No. 2017YFA0304202) and the Air Force Office of Scientific Research (Grant No. FA9550-18-1-0141).

\section*{SUPPLEMENTARY MATERIAL}

\section*{I. Solving the two-mode quantum Langevin equations in the time domain}

Here we provide the details of how to solve the QLEs Eq.~\eqref{QLE2}, which gives the analytical expressions Eq.~\eqref{SYSx} of the variance of the squeezed cavity and magnon quadratures. Equation~\eqref{QLE2} can be cast in the matrix form
\begin{equation}\label{uuAAnn}
\dot{u} (t) = A u(t) + n(t) ,
\end{equation}
where $u(t)=\big[\delta X (t), \delta Y (t), \delta x (t), \delta y (t) \big]^T$, with the quadrature fluctuation operators $\delta X=(\delta a + \delta a^{\dag})/\sqrt{2}$, $\delta Y=i(\delta a^{\dag} - \delta a)/\sqrt{2}$, $\delta x=(\delta m + \delta m^{\dag})/\sqrt{2}$, and $\delta y=i(\delta m^{\dag} - \delta m)/\sqrt{2}$, $n (t) = \big[ \!\sqrt{2\kappa_a} X^{\rm in} (t), \sqrt{2\kappa_a} Y^{\rm in} (t), \sqrt{2\kappa_m} x^{\rm in} (t), \sqrt{2\kappa_m} y^{\rm in} (t) \big]^T$ is the vector of input noises, where $ X^{\rm in}=(a^{\rm in} + a^{\rm in \dag})/\sqrt{2}$, $Y^{\rm in}=i(a^{\rm in \dag} - a^{\rm in})/\sqrt{2}$, $x^{\rm in}=(m^{\rm in} + m^{\rm in \dag})/\sqrt{2}$, and $y^{\rm in}=i(m^{\rm in \dag} - m^{\rm in})/\sqrt{2}$, and the drift matrix $A$ is given by  
\begin{equation}\label{AAA}
A =
\begin{pmatrix}
-\kappa_a  &  \Delta_a  &  0 &  g_{ma}    \\
-\Delta_a  & -\kappa_a  & -g_{ma}  & 0    \\
0 & g_{ma}  & -\kappa_m  & \Delta_m  \\
-g_{ma}  & 0 & - \Delta_m & -\kappa_m  \\
\end{pmatrix}.
\end{equation}
The formal solution of Eq.~\eqref{uuAAnn} takes the following form
\begin{equation}
u(t)=Z(t)u(0) +\!\! \int_0^t dt' Z(t') n(t-t'),
\end{equation}
where $Z(t)=e^{At}$, and $u(0)$ is the initial state of $u(t)$. We are interested in the variances of the cavity and magnon mode quadratures, which are the diagonal elements of the time-dependent covariance matrix (CM) $V(t)$, which is defined as $V_{ij} (t) = \frac{1}{2} \langle u_i(t) u_j(t') + u_j(t') u_i(t) \rangle$ ($i,j=1,2,3,4$). The CM $V(t)$ can be obtained by
\begin{equation}
V(t)=Z(t)V(0)Z^T(t) +\!\! \int_0^t dt' Z(t') D Z^T(t') ,
\end{equation}
where $V(0)$ is the CM of the initial state, and $D$ is the diffusion matrix, defined as $\langle  n_i(t) n_j(t') +n_j(t') n_i(t) \rangle/2 = D_{ij} \delta (t-t')$, representing the correlations between the input noise terms, which is in the form of
\begin{widetext}
\begin{equation}\label{DDD}
D =
\begin{pmatrix}
\kappa_a (2{\cal N}+1 + {\cal M} +{\cal M}^* )  &  i \kappa_a ({\cal M}^* - {\cal M})  &  0 &  0    \\
i \kappa_a ({\cal M}^* - {\cal M})   & \kappa_a (2{\cal N}+1 - {\cal M} - {\cal M}^* )  & 0  & 0    \\
0 & 0  & \kappa_m (2N_m+1)  & 0  \\
0 & 0 & 0 & \kappa_m (2N_m+1)  \\
\end{pmatrix}.
\end{equation}
\end{widetext}
We are particularly interested in the quadrature squeezing in the steady state. When the system is stable, $Z(\infty)=0$, the CM $V(t)$ for $t \to \infty$ becomes
\begin{equation}
V(\infty)= \int_0^{\infty} dt' Z(t') D Z^T(t').
\end{equation}
The solution of $V(\infty)$ is equivalent to the solution of the so-called Lyapunov equation~\cite{Parks}
\begin{equation}\label{Lyapunov}
AV + VA^T = -D ,
\end{equation}
which is linear and can be straightforwardly solved. For the resonant case $\Delta_a = \Delta_m = 0$, relatively simple expressions for the variances of the four mode quadratures can be obtained, in which the cavity phase and magnon amplitude quadratures are squeezed, whose variances are
\begin{widetext}
\begin{equation}
\begin{split}
\langle \delta Y(t)^2  \rangle  &=  \frac{ g_{ma}^2 (2N_m+1) \kappa_m +\kappa_a (g_{ma}^2 + \kappa_a \kappa_m +\kappa_m^2 )  (\cosh 2r -\cos \theta \sinh 2r )     }{2(\kappa_a+\kappa_m) (g_{ma}^2+\kappa_a \kappa_m) },   \\
\langle \delta x(t)^2  \rangle  &=  \frac{  (2N_m+1) \kappa_m (g_{ma}^2 + \kappa_a \kappa_m +\kappa_a^2 ) + g_{ma}^2 \kappa_a (\cosh 2r -\cos \theta \sinh 2r )     }{2(\kappa_a+\kappa_m) (g_{ma}^2+\kappa_a \kappa_m) }. 
\end{split}
\end{equation}
\end{widetext}
For the nonresonant case, $\Delta_a$, $\Delta_m \ne 0$, the expressions for the variances are, however, too lengthy to be reported.

\section*{II. Solving the three-mode quantum Langevin equations in the frequency domain}

For the study of mechanical squeezing, the QLEs of Eq.~\eqref{uAn} is given in the reference frame rotating at the magnon drive frequency $\omega_0$, which is red-detuned with respect to the magnon, cavity, and squeezed drive frequency $\Delta_m$, $\Delta_a$, $\Delta_s>0$ [see Fig.1 (b) in the main text]. In this case, two of the cavity input-noise correlation functions  $\langle a^{\rm in}(t) \, a^{\rm in}(t')\rangle$ and $\langle a^{\rm in \dag}(t) \, a^{\rm in \dag}(t')\rangle$ are time-dependent owing to a nonzero $\Delta_s$. This leads to a time-dependent diffusion matrix $D$ and thus the Lyapunov equation~\eqref{Lyapunov} cannot be readily used for the steady-state solutions. We thus solve the QLEs~\eqref{uAn} in the frequency domain by taking the Fourier transform of each equation. The expressions for the cavity, magnon, and mechanical quadrature fluctuations, $\delta Q (\omega)$ ($Q=X,Y,x,y,q,p$), can be obtained, which take the form of
\begin{equation}
\begin{split}
\delta Q (\omega)&= Q_A (\omega) \, a^{\rm in}(\omega) +Q_B (\omega) \, a^{\rm in \dag}(-\omega) \\
&+ Q_C (\omega) \, m^{\rm in}(\omega) +Q_D (\omega) \, m^{\rm in \dag}(-\omega) +  Q_E (\omega) \, \xi (\omega),
\end{split}
\end{equation}
where $Q_j (\omega)$ ($j=A,B,C,D,E$) are the coefficients associated with different input noises. By using the input noise correlations in the frequency domain 
\begin{equation}\label{CFfd}
\begin{split}
\langle a^{\rm in}(\omega) \, a^{\rm in \dag}(-\Omega)\rangle &= 2\pi \,({\cal N}{+}1) \, \delta(\omega+\Omega),   \\
\langle a^{\rm in \dag}(-\omega) \, a^{\rm in}(\Omega)\rangle &= 2\pi \, {\cal N} \, \delta(\omega+\Omega),   \\
\langle a^{\rm in}(\omega) \, a^{\rm in}(\Omega)\rangle &= 2\pi \, {\cal M} \, \delta(\omega+\Omega - 2 \Delta_s),   \\ 
\langle a^{\rm in \dag}(-\omega) \, a^{\rm in \dag}(-\Omega)\rangle &= 2\pi \, {\cal M}^* \, \delta(\omega+\Omega + 2 \Delta_s),  \\
\langle m^{\rm in}(\omega) \, m^{\rm in \dag}(-\Omega)\rangle &= 2\pi \, (N_m{+}1) \, \delta(\omega+\Omega),    \\
\langle m^{\rm in \dag}(-\omega) \, m^{\rm in}(\Omega)\rangle &= 2\pi \, N_m \, \delta(\omega+\Omega),    \\
\langle \xi(\omega)\xi(\Omega) \rangle  &\simeq 2\pi \, \gamma_b (2 N_b {+}1) \, \delta(\omega+\Omega),
\end{split}
\end{equation}
the spectra of fluctuations in the quadratures of the mechanical mode can be achieved by
\begin{equation}
S_Q(\omega) = \frac{1}{4\pi} \! \int_{-\infty}^{+\infty} \!\!\! d\Omega \, e^{-i (\omega+\Omega)t}  \langle \delta Q(\omega) \delta Q(\Omega) {+} \delta Q(\Omega) \delta Q(\omega) \rangle,
\end{equation}
where $Q=q,p$. The variances of the amplitude and phase quadratures of the mechanical mode are defined as
\begin{equation}
\langle \delta Q(t)^2  \rangle = \frac{1}{2\pi} \! \int_{-\infty}^{+\infty} \!\! d\omega \, S_Q(\omega) .
\end{equation}
The variance of the squeezed quadrature $\langle \delta q(t)^2  \rangle$ can be written in the following form (setting $ \Delta_s = \omega_b$)
\begin{equation}\label{c0c1}
\langle \delta q(t)^2  \rangle = \frac{1}{2\pi} \! \int_{-\infty}^{+\infty} \!\! \,  \omega_b^2 \, \bigg\{ A(\omega)  +\Big[  B(\omega)  e^{-i2\omega_b t} +{\rm c.c.}   \Big]   \bigg\}  \, d\omega ,
\end{equation}
where $A(\omega)$ and $B(\omega)$ are too lengthy to be reported. The explicit time dependence in Eq.~\eqref{c0c1} can be eliminated by working in the interaction picture with respect to the free Hamiltonian of the mechanical oscillator~\cite{GA09}, and thus we obtain
\begin{equation}
\begin{split}
\,\,\,\,    \langle \delta \tilde{q}^2  \rangle = & \frac{1}{2\pi} \!  \int_{-\infty}^{+\infty} \!\! \,  \omega_b^2 \,  \Bigg\{  \frac{1}{2} \Big(1 + \frac{\omega^2}{\omega_b^2} \Big) A(\omega)    \\
&+\bigg[  \frac{1}{4} \Big(1 + \frac{\omega}{\omega_b} \Big) \Big( 3 - \frac{\omega}{\omega_b} \Big) B(\omega)  +{\rm c.c.}   \bigg]   \Bigg\}  \, d\omega .
\end{split}
\end{equation}

\section*{III. Observing the squeezing spectrum of the cavity output to infer the magnon squeezing}

Here we provide a straightforward way to infer the squeezed state of the magnon mode. This is realized by directly measuring the spectrum of the cavity output field (see Fig.1 in the main text) and by observing the strong coupling effects of the cavity and magnon modes. Owing to the intrinsic beamsplitter (state-swap) interaction between the cavity and the magnons, the squeezed cavity field and the cavity-magnon strong coupling are sufficient to infer that the magnon mode is prepared in a squeezed state.

\begin{figure*}[t]
\hskip-0.4cm\includegraphics[width=0.85\linewidth]{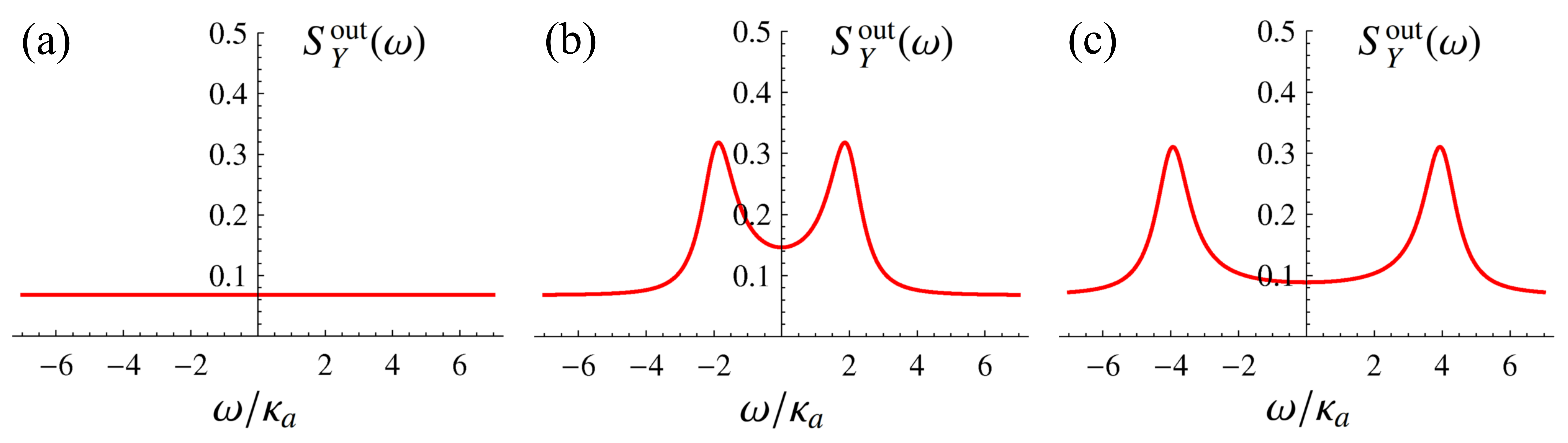} 
\caption{The spectrum $S_Y^{\rm out} (\omega)$ of the phase fluctuation of the output field versus the frequency $\omega$ for (a) $g_{ma}=0$, (b) $g_{ma}=2 \kappa_a$, and (c) $g_{ma}=4 \kappa_a$. We take $r=1$ and the other parameters are as in Fig. 3 (a) and (b) in the main text.  }
\label{figSM}
\end{figure*}

We thus calculate the spectrum of the cavity output field. The fluctuation of the cavity output field $\delta a^{\rm out} (\omega)$ can be obtained by using the standard input-output formula~\cite{Gardiner84}
\begin{equation}
\delta a^{\rm out} = \sqrt{2 \kappa_a} \delta a - a^{\rm in}, 
\end{equation}
where $\delta a (\omega)$ can be easily achieved by solving the QLEs Eq.~\eqref{QLE2} in the frequency domain. This allows us to define the general quadrature fluctuation of the output field as 
\begin{equation}
\delta Z^{\rm out} (\omega) = \frac{1}{\sqrt{2}} \Big[ \delta a^{\rm out} (\omega) e^{-i \phi} + \delta a^{\rm out \dag} (-\omega) e^{i \phi}  \Big],
\end{equation}
with $\phi$ being the phase. For $\phi=0$, $\delta Z^{\rm out} (\omega) = \delta X^{\rm out} (\omega) $, which is the amplitude fluctuation of the output field; for $\phi=\frac{\pi}{2}$, $\delta Z^{\rm out} (\omega) = \delta Y^{\rm out} (\omega) $, i.e., the phase fluctuation of the output field. Through straightforward calculations, $\delta Z^{\rm out} (\omega)$ is found to be in the following form
\begin{equation}
\begin{split}
\delta Z^{\rm out} (\omega) &= {\cal A} (\omega)\, a^{\rm in}(\omega) +  {\cal B} (\omega) \, a^{\rm in \dag}(-\omega) \\
&\,+  {\cal C} (\omega) \, m^{\rm in}(\omega) +  {\cal D} (\omega) \, m^{\rm in \dag}(-\omega) ,
\end{split}
\end{equation}
where the coefficients ${\cal O} (\omega) $ (${\cal O}={\cal A},{\cal B},{\cal C},{\cal D}$) are given by
\begin{equation}
\begin{split}
{\cal A} (\omega) &= \frac{e^{-i \phi}}{\sqrt{2}} \, \Bigg[ -1 + \frac{ 2 i \kappa_a (\Delta m - i \kappa_m -\omega )}{ g^2_{ma} - (\Delta a - i \kappa_a -\omega) (\Delta m - i \kappa_m -\omega ) }    \Bigg],   \\
{\cal B} (\omega) &= \frac{e^{i \phi}}{\sqrt{2}} \, \Bigg[ -1 - \frac{ 2 i \kappa_a (\Delta m + i \kappa_m +\omega )}{ g^2_{ma} - (\Delta a + i \kappa_a +\omega) (\Delta m + i \kappa_m +\omega ) }    \Bigg],   \\
{\cal C} (\omega) &= - e^{-i \phi} \frac{ i g_{ma} \sqrt{ 2 \kappa_a \kappa_m } }{ g^2_{ma} - (\Delta a - i \kappa_a -\omega) (\Delta m - i \kappa_m -\omega )  },   \\
{\cal D} (\omega) &=  e^{i \phi}  \frac{ i g_{ma} \sqrt{ 2 \kappa_a \kappa_m } }{  g^2_{ma} - (\Delta a + i \kappa_a +\omega) (\Delta m + i \kappa_m +\omega )  }.   \\
\end{split}
\end{equation}
The spectrum of the quadrature fluctuation $\delta Z^{\rm out} (\omega)$ of the output field is defined by
\begin{equation}
\begin{split}
&S_Z^{\rm out} (\omega) =  \\
&\frac{1}{4\pi} \! \int_{-\infty}^{+\infty} \!\!\! d\Omega \, e^{-i (\omega+\Omega) t } \, \langle \delta Z^{\rm out} (\omega) \delta Z^{\rm out} (\Omega)  +  \delta Z^{\rm out} (\Omega) \delta Z^{\rm out} (\omega) \rangle  .
\end{split}
\end{equation}
By using the input noise correlations in Eq.~\eqref{CFfd}, the spectrum $S_Z^{\rm out} (\omega)$ can be obtained, which is, however, too lengthy to be reported here. The output field is squeezed if $S_Z^{\rm out} (\omega)$ is smaller than that of the vacuum state, i.e., $S_Z^{\rm out} (\omega)<\frac{1}{2}$. After exploring the full range of the phase $\phi \in [0,2\pi)$, we find that the phase quadrature ($\phi = \frac{\pi}{2}$) of the output field is most squeezed, and we plot its spectrum $S_Y^{\rm out} (\omega)$ in Fig.~\ref{figSM} using the parameters as in Fig. 3 (a) and (b) in the main text but for $r=1$ and $g_{ma}=0$, $2\kappa_a$, and $4\kappa_a$. We observe two peaks in the spectrum at $\omega \simeq \pm g_{ma}$ when the coupling is large $g_{ma}>\kappa_a$, which is a clear signature of the strong coupling between the cavity and magnon modes. The magnon mode is therefore squeezed due to this strong coupling and the transfer of squeezing from the cavity field to the magnons is most pronounced at frequencies $\omega \simeq g_{ma}$, and $\omega \simeq -g_{ma}$.


\begin{thebibliography}{99}

\bibitem{Kittel}
C. Kittel, Phys. Rev. {\bf 73}, 155 (1948).

\bibitem{Strong1}
H. Huebl {\it et al}., Phys. Rev. Lett. {\bf 111}, 127003 (2013). 
\bibitem{Strong2}
Y. Tabuchi {\it et al}., Phys. Rev. Lett. {\bf 113}, 083603 (2014). 
\bibitem{Strong3}
X. Zhang {\it et al}., Phys. Rev. Lett. {\bf 113}, 156401 (2014). 
\bibitem{Strong4}
M. Goryachev {\it et al}., Phys. Rev. Appl. {\bf 2}, 054002 (2014). 
\bibitem{Strong5}
L. Bai {\it et al}., Phys. Rev. Lett. {\bf 114}, 227201 (2015). 
\bibitem{Strong6}
D. Zhang {\it et al.} npj Quantum Information {\bf 1}, 15014 (2015).


\bibitem{TangNC}
X. Zhang {\it et al.}, Nat. Commun. {\bf 6}, 8914 (2015).

\bibitem{YouNC}
D. Zhang, X.-Q. Luo, Y.-P. Wang, T.-F. Li, and J. Q. You, Nat. Commun. {\bf 8}, 1368 (2017).

\bibitem{spinCur}
L. Bai {\it et al.}, Phys. Rev. Lett. {\bf 118}, 217201 (2017).

\bibitem{You18}
Y.-P. Wang {\it et al.}, Phys. Rev. Lett. {\bf 120}, 057202 (2018).


\bibitem{Naka15}
Y. Tabuchi {\it et al.}, Science {\bf 349}, 405 (2015).

\bibitem{Tang16}
X. Zhang, C.-L. Zou, L. Jiang, and H. X. Tang, Sci. Adv. {\bf 2}, e1501286 (2016).

\bibitem{Naka17}
D. Lachance-Quirion {\it et al}., Sci. Adv. {\bf 3}, e1603150 (2017).

\bibitem{JiePRL}
J. Li, S.-Y. Zhu, and G. S. Agarwal, Phys. Rev. Lett. {\bf 121}, 203601 (2018).


\bibitem{Caves}
C. M. Caves, K. S. Thorne, R. W. P. Drever, V. D. Sandberg, and M. Zimmermann, Rev. Mod. Phys. {\bf 52}, 341 (1980).

\bibitem{collapse}
A. Bassi, K. Lochan, S. Satin, T. P. Singh, and H. Ulbricht, Rev. Mod. Phys. {\bf 85}, 471 (2013).

\bibitem{Loock}
S. L. Braunstein and P. van Loock, Rev. Mod. Phys. {\bf 77}, 513 (2005).

\bibitem{OMRMP}
M. Aspelmeyer, T. J. Kippenberg, and F. Marquardt, Rev. Mod. Phys. {\bf 86}, 1391 (2014).

\bibitem{Schwab}
E. E. Wollman {\it et al}., Science {\bf 349}, 952 (2015).

\bibitem{Simon16}
R. Riedinger {\it et al}., Nature (London) {\bf 530}, 313 (2016).

\bibitem{enMM1}
R. Riedinger {\it et al}., Nature (London) {\bf 556}, 473 (2018).

\bibitem{enMM2}
C. F. Ockeloen-Korppi {\it et al}., Nature (London) {\bf 556}, 478 (2018).



\bibitem{Kumar}
J. E. Sharping, M. Fiorentino, and P. Kumar, Opt. Lett. {\bf 26}, 367 (2001).

\bibitem{Zoller}
K. J\"ahne, C. Genes, K. Hammerer, M. Wallquist, E. S. Polzik, and P. Zoller, Phys. Rev. A {\bf 79}, 063819 (2009).

\bibitem{GA09}
S. Huang and G. S. Agarwal, New J. Phys. {\bf 11}, 103044 (2009).


\bibitem{Kittel2}
C. Kittel, Phys. Rev. {\bf 110}, 836 (1958).

\bibitem{You16}
Y.-P. Wang {\it et al}., Phys. Rev. B {\bf 94}, 224410 (2016).

\bibitem{HPT}
T. Holstein and H. Primakoff, Phys. Rev. {\bf 58}, 1098 (1940).



\bibitem{Gardiner}
%C. W. Gardiner and P. Zoller, {\it Quantum Noise} (Springer, Berlin, Germany, 2000).
C. W. Gardiner, Phys. Rev. Lett. {\bf 56}, 1917 (1986).

\bibitem{JPA1}
B. Yurke, J. Opt. Soc. Am. B {\bf 4}, 1551 (1987); B. Yurke {\it et al.}, Phys. Rev. Lett. {\bf 60}, 764 (1988); B. Yurke {\it et al.}, Phys. Rev. A {\bf 39}, 2519 (1989).
\bibitem{JPA2}
R. Movshovich {\it et al.}, Phys. Rev. Lett. {\bf 65}, 1419 (1990).
\bibitem{JPA3}
M. A. Castellanos-Beltran, K. D. Irwin, G. C. Hilton, L. R. Vale, and K. W. Lehnert, Nat. Phys. {\bf 4}, 929 (2008).
\bibitem{JPA4}
T. Yamamoto {\it et al.}, Appl. Phys. Lett. {\bf 93}, 042510 (2008).
\bibitem{JPA5}
F. Mallet {\it et al.}, Phys. Rev. Lett. {\bf 106}, 220502 (2011).
\bibitem{JPA6}
E. P. Menzel {\it et al.}, Phys. Rev. Lett. {\bf 109}, 250502 (2012).
\bibitem{JPA7}
L. Zhong {\it et al.}, New J. Phys. {\bf 15}, 125013 (2013).
\bibitem{JPA8}
K. G. Fedorov {\it et al.}, Phys. Rev. Lett. {\bf 117}, 020502 (2016).
\bibitem{JPA9}
S. Kono {\it et al.}, Phys. Rev. Lett. {\bf 119}, 023602 (2017).
\bibitem{JPA10}
A. Bienfait {\it et al.}, Phys. Rev. X {\bf 7}, 041011 (2017).
\bibitem{JPA11}
M. Malnou, D. A. Palken, Leila R. Vale, Gene C. Hilton, and K. W. Lehnert, Phys. Rev. Applied {\bf 9}, 044023 (2018).


\bibitem{Siddiqi}
K. W. Murch, S. J. Weber, K. M. Beck, E. Ginossar, and I. Siddiqi, Nature {\bf 499}, 62 (2013).


\bibitem{Markov}
V. Giovannetti and D. Vitali, Phys. Rev. A {\bf 63}, 023812 (2001); R. Benguria and M. Kac, Phys. Rev. Lett. {\bf 46}, 1 (1981).

\bibitem{Note}
In the experiment of Ref.~\cite{Tang16} a small value of the magnon-phonon coupling $g_{mb}/2\pi<10^{-2}$ Hz has been measured, while the magnon-cavity coupling $g_{ma}$ can be 8 or 9 orders of magnitude larger than $g_{mb}$, as reported in Refs.~\cite{Strong1,Strong2,Strong3,Strong4,Strong5}.

\bibitem{SM}
See Supplemental Material for the details of solving the QLEs in the two cases of magnon and mechanical squeezing, and for the squeezing spectrum of the cavity output for inferring the magnon squeezing, which includes Refs.~\cite{GA09,Parks,Gardiner84}.

\bibitem{Parks}
P. C. Parks and V. Hahn, {\it Stability Theory} (Prentice Hall, New York, U.S., 1993).

\bibitem{Gardiner84}
M. J. Collett and C. W. Gardiner, Phys. Rev. A {\bf 30}, 1386 (1984).


\bibitem{Note3}
Here we adopt slightly different parameters from those (of Ref.~\cite{Strong2}) in the study of magnon squeezing. One may choose the parameters according to the purpose of producing squeezed states of magnons or phonons.   


\bibitem{Jie18}
J. Li, S. Gr\"oblacher, S.-Y. Zhu, and G. S. Agarwal, Phys. Rev. A {\bf 98}, 011801(R) (2018). 

\bibitem{Hofer}
S. G. Hofer, W. Wieczorek, M. Aspelmeyer, and K. Hammerer, Phys. Rev. A {\bf 84}, 052327 (2011).

\bibitem{B0P}
The relation between the drive magnetic field $B_0$ and the power $P$ is $B_0=\frac{1}{R} \sqrt{\frac{2P\mu_0}{\pi c}}$~\cite{JiePRL}, with $R$ the radius of the YIG sphere, $c$ the speed of an electromagnetic wave propagating through the vacuum, and $\mu_0$ the vacuum magnetic permeability.

\bibitem{note2}
The magnon mode (spin wave) is a collective mode, which contains a large number of spins ($3.5 \times 10^{16}$ for a 250-$\mu$m-diameter YIG sphere), and we consider that it is at macroscopic scale and can be referred to as a macroscopic quantum state.










\end{thebibliography}
\end{document}